\documentstyle[preprint,aps]{revtex}

\begin{document}

\title {Quantization of the Taub cosmological model with
extrinsic time}

\author{Gabriel Catren{\footnote{Electronic address:
gabrielcatren@hotmail.com}}}

\address{
{\it Instituto de Astronom\'\i a y F\'\i sica del Espacio, Casilla de Correo 67 - Sucursal
28, 1428 Buenos Aires, Argentina\\}}

\author{Rafael Ferraro{\footnote{Electronic address:
ferraro@iafe.uba.ar}}}

\address{
{\it Instituto de Astronom\'\i a y F\'\i sica del Espacio, Casilla de Correo 67 - Sucursal
28, 1428 Buenos Aires, Argentina\\ and Departamento de F\'\i sica, Facultad de Ciencias
Exactas y Naturales, Universidad de Buenos Aires - Ciudad Universitaria, Pabell\' on I,
1428 Buenos Aires, Argentina\\}}

\maketitle

\begin{abstract}

{\tighten The paper addresses the quantization of minisuperspace cosmological models, with
application to the Taub Model. By desparametrizing the model with an extrinsic time, a
formalism is developed in order to define a conserved Schr\"{o}dinger inner product in the
space of solutions of the Wheeler-De Witt equation. A quantum version of classical
canonical transformations is introduced for connecting the solutions of the Wheeler-De
Witt equation with the wave functions of the desparametrized system. Once this
correspondence is established, boundary conditions on the space of solutions of the
Wheeler-De Witt equation are found to select the physical subspace. The question of
defining boundary conditions on the space of solutions of the Wheeler-De Witt equation
without having reduced the system is examined.

}

\end{abstract}
\vskip  1cm

PACS numbers: 98.80.Hw, 04.20.Cv

\narrowtext

\newpage

\section{Introduction}

\smallskip General Relativity is an example of parametrized system, i.e. a
system whose action is invariant under changes of the integrating parameter $%
\tau $ (``reparametrization''), this invariance being a consequence of the covariance of
the theory. This means that in General Relativity there is no privileged time variable. On
the contrary, in the ordinary formulation of quantum mechanics there is a time parameter
besides the true degrees of freedom, and the inner product remains conserved in the time
evolution of the system. This difference between General Relativity and quantum mechanics,
known as the problem of time \cite{canada,isham,extrinseco}, is one of the main obstacles
for finding a quantum theory of gravity.\smallskip

The evolution of a dynamical system is characterized by the way in which its
dynamical variables evolve as a function of time. In this formulation time
is a relevant physical parameter clearly distinct from the dynamical
variables. There is nevertheless an alternative formulation of dynamics
(parametrized systems) in which time is mixed with the dynamical variables
\cite{lanczos}. A parametrized system can be obtained from an action $%
S\left( q^{\mu },p_{\mu }\right) $ which is not invariant under
reparametrizations by raising the time to the rank of a dynamical variable.
Let us start with an action of the form
\begin{mathletters}
\[
S\left[ q^{\mu },p^{\mu }\right] =\int_{t_{1}}^{t_{2}}p_{\mu }dq^{\mu
}-h(q^{\mu },p_{\mu },t)dt\text{ }\mu =1,....,n
\]
By identifying $q^{0}\equiv t,$ $p_{0}\equiv -h$ one can rewrite the
integrand as $p_{i}dq^{i}=p_{i}(dq^{i}/d\tau )d\tau $, $i=0,.....,n.$ In
this way the extended set of variables are left as functions of some
physically irrelevant parameter $\tau .$ The set $\left\{
q^{i},p_{i}\right\} $ can be independently varied provided that the
definition of $p_{t}$ is incorporated to the action as a constraint $%
H=p_{0}+h(q^{\mu },p_{\mu },t)=0,$ so yielding the following action
\end{mathletters}
\begin{equation}
S[q^{i}(\tau ),p_{i}(\tau ),N(\tau )]=\int_{\tau _{1}}^{\tau _{2}}\left(
p_{i}\frac{dq^{i}}{d\tau }-NH(q,p)\right) d\tau  \label{iu}
\end{equation}
where $N$ is the Lagrange multiplier whose variation assures that the
constraint does hold. This action is invariant under reparametrizations $%
\tau \rightarrow \tau +\varepsilon (\tau )$. The time variable $t$ satisfies
the Poisson bracket
\begin{equation}
\left\{ t,H\right\} =1  \label{po}
\end{equation}

Once the system has been parametrized, it can be reduced using any time
variable provided that it satisfies $\left( \ref{po}\right) $. This kind of
time variables are called global times $\cite{tiempo}$. In order to
generalize this restriction let us suppose that we know a globally well
defined time variable $\tilde{t}=\tilde{t}\left( q^{i},p_{i}\right) $which
satisfies
\begin{equation}
\left\{ \tilde{t},H\right\} \mid _{H=0}=f\left( q,p\right) >0  \label{f}
\end{equation}

The important fact is that $f$ has a definite sign on the constraint surface
(it could also be negative). In this case the variable $\tilde{t}$ is a
global time associated with the Hamiltonian $\tilde{H}\equiv $ $f^{-1}\left(
q,p\right) H.$

The constraint $\tilde{H}=0$ could also be expressed in a set of variables
in which the Hamiltonian $H$ has not the form $H=p_{0}+h$. In fact we can
perform a canonical transformation
\[
\left\{ q^{i},p_{i}\right\} =\left\{ q^{o}=t,p_{0}=-h,q^{\mu },p_{\mu
}\right\} \rightarrow \left\{ Q^{i},P_{i}\right\}
\]
where now the time is hidden among the rest of the variables. In other
words, a constraint of the form $H=p_{0}+h$ can be disguised by scaling it
or by performing canonical transformations.

One of the main properties of the Hamiltonian formulation of General\
Relativity is that the Hamiltonian is constrained to be zero, making
manifest that General Relativity is a parametrized system. In cases like
this one, in which the theory is an already parametrized system, the
invariance under reparametrizations means that there is no privileged time
variable. To reduce the system means to select among the dynamical variables
a proper global time, i.e., a variable which monotonically increases along
any dynamical trajectory, to work as a physical clock. In this way we can
express the evolution of the canonical variables as a function of this
physical clock. The first step to reduce the system is thus to perform a
canonical transformation in order to find a set of variables $\left\{
q^{i},p_{i}\right\} $ where the variables $q_{0}=t$ is a global time. The
Hamilton equations are
\begin{eqnarray*}
\frac{dt}{d\tau } &=&Nf \\
\frac{dq_{\mu }}{d\tau } &=&Nf\frac{\partial h}{\partial p^{\mu }} \\
\frac{dp^{\mu }}{d\tau } &=&-Nf\frac{\partial h}{\partial q_{\mu }}
\end{eqnarray*}

The dynamics of the system is thus undetermined unless one fixes a gauge,
i.e., unless one chooses a physical clock. Choosing the gauge $\tau =t$
means choosing $N\left( \tau \right) =\frac{1}{f\left[ q_{\mu }\left( \tau
\right) ,p^{\mu }\left( \tau \right) \right] }$.

One of the mains approximations for quantizing General Relativity begins by
reformulating it under a Hamiltonian formulation (ADM formalism $\cite
{gravitation}$) . Within the framework of this canonical formalism or
geometrodynamics it is supposed that the Lorentzian space-time manifolds $M$
are diffeomorphic to $R\times S$ where $S$ represents a collection of
spacelike hypersurfaces $\Sigma $ parametrized by a real time parameter $t$
(foliation). The Riemannian metric $g_{ij}$ of one of these hypersurfaces $%
\Sigma $ play the role of the configuration variable. The analogous of the
configuration space $R^{n}$ is the space of all the Riemannian metrics $%
g_{ij}$ called superspace. The conjugate momentum $\pi ^{ij}$ is directly
related with the way in which the hypersurface $\Sigma $ is embedded in the
manifold $M,$ i.e., with the extrinsic curvature of the hypersurface $\Sigma
$. The covariance of the theory under general coordinate transformations is
reflected within this formalism in the presence of four constraints per each
point of space-time. The so called Hamiltonian constraint assures the
invariance of the theory under a changing of the foliation, while the
momentum constraints assure the invariance under a change of the spatial
coordinates used to represent the spatial geometry of each hypersurface. The
states of the corresponding quantum theory $\Psi \left[ g_{ij}\right] $ are
functionals of the spatial metric $g_{ij}$ which satisfies the quantum
version of the classical constraints in accordance with the Dirac method.
The quantum version of the momentum constraints implies that the wave
function depends on the geometry $^{3}g$ of the hypersurface but not on the
particular metric tensor $g_{ij}$ used to represent it. The quantum version
of the Hamiltonian constraint is the so called Wheeler-De Witt equation.

Many of the tentatives for quantizing General Relativity began addressing
the analogy between the Wheeler-De Witt equation and the Klein-Gordon
equation. In fact both systems have Hamiltonian constraints which are
hyperbolic in the momenta. The constraint associated with the motion of a
particle in a pseudo-Riemannian geometry has the form
\begin{equation}
H_{particle}=g^{ij}\left( q^{k}\right) p_{i}\text{ }p_{j}-m^{2}=0
\label{hampart}
\end{equation}
The space of solutions of the Klein-Gordon equation can be turned into a
Hilbert space with a positive definite inner product only if the background
is stationary. In this case the Hilbert space of the physical states will be
the subspace of positive norm, this being equivalent to consider just one of
the sheets of the hyperbolic constraint surface. Choosing the coordinates in
a way that $g^{\mu 0}=0$ $\left( \Rightarrow g^{00}=g_{00}^{-1}\right) $ and
calling $\gamma ^{\mu \nu }\equiv -g_{00}g^{\mu \nu }$ we can write $\left(
\ref{hampart}\right) $ in the form
\begin{eqnarray}
H_{particle} &=&g^{00}\left( p_{0}p_{0}-\gamma ^{\mu \nu }p_{\mu }p_{\nu
}-g_{00}m^{2}\right)  \label{o} \\
&=&g^{00}\left( p_{0}-\sqrt{\gamma ^{\mu \nu }p_{\mu }p_{\nu }+g_{00}m^{2}}%
\right) \left( p_{0}+\sqrt{\gamma ^{\mu \nu }p_{\mu }p_{\nu }+g_{00}m^{2}}%
\right)  \nonumber
\end{eqnarray}
In addition it is necessary to find a temporal Killing vector of the
supermetric which also should be a symmetry of the potential term (this
property could be relaxed to a conformal Killing vector). In this case the
proper time variable to reduce the system is the parameter of the Killing
vector. Otherwise there would be pair creation. In order to build a good
analogy with the relativistic particle is also necessary to have a positive
definite potential term for playing the role of the mass term. If the
potential is positive definite, the momentum $p_{0}$ does not go to zero on
the constraint surface $H=0.$ This means that the Poisson bracket $%
\{q_{0},H\}=2g^{00}p_{0}$ has a definite sign on each sheet of the
constraint surface. If we choose $p_{0}+\sqrt{\gamma ^{\mu \nu }p_{\mu
}p_{\nu }+g_{00}m^{2}}=0,$ the momentum $p_{o},$ and so $\{q_{0},H\},$ will
be negative on this sheet (provided that $g^{00}>0)$. The other factor has
then a definite sign on this sheet playing the role of the function $f$
defined in $\left( \ref{f}\right) .$ In this case $f$ will be negative,
being this the reason why $\{q_{0}=t,H\}<0.$ But $t$ is still the variable
which monotonically increases on any dynamical trajectory because $\left\{ t,%
\tilde{H}\right\} =1$ where $\tilde{H}=\frac{H}{p_{0}-\sqrt{\gamma ^{\mu \nu
}p_{\mu }p_{\nu }+g_{00}m^{2}}}.$ The quantum physical states can be
obtained by solving a Schr\"{o}dinger equation with a positive definite
operator $\hat{h}$ associated with the Hamiltonian of the reduced system $%
\sqrt{\gamma ^{\mu \nu }p_{\mu }p_{\nu }+g_{00}m^{2}}.$ As it was said
before, to fix the gauge $t=\tau $ implies to choose $N=\frac{1}{f}$. The
relation between proper time and the time variable chosen to represent the
hypersurfaces of simultaneity is $dT=$ $f^{-1}dt$ where $T$ is the proper
time, this being a consequence of the way in which the space-time interval
is expressed in the ADM formalism. In geometrodynamics there is a conformal
Killing vector of the supermetric \cite{killing} but this vector is not as
well a symmetry of the potential term. Besides this potential term is the
spatial curvature \cite{gravitation}, which can be negative in some regions
of the configuration space. Thus it is not possible to associate an operator
$\hat{h}$ with the square root, as one effectively does for the relativist
particle.

There are thus two main approaches for achieving this quantization program.
One possibility is to quantize the system without reducing it. The resulting
Wheeler-De Witt equation is an hyperbolic equation while the Schr\"{o}dinger
equation associated with the reduced system is a parabolic one. The former
has then more solutions than the latter, so being necessary to define
boundary conditions in order to select the physical solutions. Besides it is
not clear in this approach how to define a conserved inner product without
having reduced the system, i.e., without knowing which variable plays the
role of time. Another proposal is to perform a canonical transformation in
order to find a Hamiltonian of the form $\left( \ref{hampart}\right) $ with
a positive definite potential term independent of the variable $q_{0}$. The
formalism of the relativistic particle can then be applied using $q_{0}$ as
a proper time variable. The system can thus be quantized by means of the
corresponding Schr\"{o}dinger equation associated with one of the sheets of
the constraint surface. The Hilbert space of the quantum states can be
endowed with the natural inner product associated with the Schr\"{o}dinger
equation. This approach has the problem that different choices of global
time variables can lead to different quantum theories.

\smallskip

\section{Proposed formalism}

\smallskip

In this work we will address the quantization of minisupersapace
cosmological models. The quantization program will be as follows. We will
start with a Hamiltonian constraint such that none of the variables is a
global time. We will suppose that it is possible to perform a coordinate
transformation so that a subsystem depending on just one pair of canonical
variables $\{q_{1},p_{1}\}$ is separated in the Hamiltonian constraint
\begin{equation}
H_{q}=h_{q_{1}}\left( q_{1},p_{1}\right) -h_{q_{\mu }}\left( q_{\mu },p_{\mu
}\right) \mu =2,..,N  \label{g}
\end{equation}
where the Hamiltonian $h_{q_{1}}\left( q_{1},p_{1}\right) $ has the form
\begin{equation}
h_{q_{1}}\left( q_{1},p_{1}\right) =\frac{4}{\left[ \frac{d\ln V}{dq_{1}}%
\right] ^{2}}\text{ }p_{1}^{2}+V\left( q_{1}\right)  \label{er}
\end{equation}
with $V\left( q_{1}\right) >0$ and $h_{q_{\mu }}>0.$ In order to find a
global time, one should look for another canonical transformation $\left\{
q_{i},p_{i}\right\} \rightarrow \left\{ Q_{i},P_{i}\right\} $ $i=1,...,N$
such that
\begin{eqnarray}
Q_{1} &=&t=t\left( q_{1},p_{1}\right)  \label{uy} \\
P_{1} &=&p_{t}=\left[ h_{q_{1}}\left( q_{1},p_{1}\right) \right] ^{\frac{1}{2%
}}  \nonumber \\
Q_{\mu } &=&q_{\mu }  \nonumber \\
P_{\mu } &=&\text{ }p_{\mu }  \nonumber \\
\mu &=&2,...,N  \nonumber
\end{eqnarray}
In this way we could separate an extrinsic time $t$, i.e., a global time
variable which is a function of both original canonical coordinates and
momentum. The generator function for the canonical transformation and the
corresponding momenta are
\begin{eqnarray}
F_{1}\left( q_{1},t\right) &=&-\sinh \left( t\right) \left[ V\left(
q_{1}\right) \right] ^{\frac{1}{2}}  \label{can} \\
p_{t} &=&\frac{\partial F_{1}}{\partial t}=-\cosh t\left[ V\left(
q_{1}\right) \right] ^{\frac{1}{2}}  \nonumber \\
p_{q_{1}} &=&\frac{\partial F_{1}}{\partial q_{1}}=-\frac{1}{2}\sinh t\left[
V\left( q_{1}\right) \right] ^{-\frac{1}{2}}\frac{dV}{dq_{1}}=-\frac{1}{2}%
\sinh t\left[ V\left( q_{1}\right) \right] ^{\frac{1}{2}}\frac{d\left( \ln
V\right) }{dq_{1}}  \nonumber
\end{eqnarray}
so that
\[
\frac{4}{\left[ \frac{d\ln V}{dq_{1}}\right] ^{2}}p_{q_{1}}^{2}+V\left(
q_{1}\right) =V\left( q_{1}\right) \sinh ^{2}t+V\left( q_{1}\right) =V\left(
q_{1}\right) \cosh ^{2}t=p_{t}^{2}
\]
The Hamiltonian in the new set of variables has the form
\begin{equation}
H_{Q}=p_{t}^{2}-h_{q_{\mu }}\left( q_{\mu },p_{\mu }\right)  \label{t}
\end{equation}
Thus, when the Hamiltonian $h_{q_{\mu }}\left( q_{\mu },p_{\mu }\right) $ is
positive definite and independent of time $t$ one gets a constraint such
that the analogy with the relativistic particle does hold, and one can
quantize the reduced model by means of the parabolic Schr\"{o}dinger
equation associated with one of the sheets of the constraint surface.

Once the system was reduced and quantized we want to find out what kind of
boundary conditions should be imposed on the solutions of the Wheeler-De
Witt equation expressed in the original set of variables. As was pointed
out, the hyperbolic Wheeler-De Witt has twice the number of independent
solutions than the parabolic Schr\"{o}dinger equation. It is then necessary
to impose proper boundary conditions for selecting the physical solutions.
In order to do that we will follow the lines of work used in \cite{beluardi}
. Knowing the classical canonical transformation for reducing the model, its
analogue in the quantum level can be defined. In Ref. \cite{gandur} the
conditions for relating the wave functions corresponding to a pair of
quantum-mechanical systems whose classical Hamiltonians are canonically
equivalent are studied. If one has two arbitrary Hamiltonians related at the
classical level by a canonical transformation corresponding to the
generating function $F_{1}\left( q,Q\right) ,$ the main issue is to find out
what kind of integral transforms can be defined in order to relate the wave
functions corresponding to each quantum-mechanical system. Generalizing the
Fourier transform, a relationship of the following kind is proposed
\begin{equation}
\Theta _{E}\left( q\right) =N\left( E\right) \int_{-\infty }^{+\infty
}dQe^{iF\left( q,Q\right) }\Phi _{E}\left( Q\right)  \label{int}
\end{equation}
where $F\left( q,Q\right) $ is not in general the generating function $%
F_{1}\left( q,Q\right) $ for the classical canonical transformation. In Ref.
$\cite{gandur}$ it is shown, however, that this function coincides in fact
with the generating function for the classical canonical transformation when
the Hamiltonians operators satisfy the condition
\begin{equation}
H_{q}\left( -i\frac{\partial }{\partial q},q\right) e^{iF\left( q,Q\right)
}=H_{Q}\left( i\frac{\partial }{\partial Q},Q\right) e^{iF\left( q,Q\right) }
\label{con}
\end{equation}
where some proper boundary conditions in the integration limits are also
assumed. If the canonical transformation cannot be represented by means of a
generating function of the first kind, analogous integral transforms and
conditions can be defined using the corresponding generating function. The
inverse of the integral transform $\left( \ref{int}\right) $ is
\begin{equation}
\Phi _{E}\left( Q\right) =N\int_{-\infty }^{+\infty }dq\left| \frac{\partial
^{2}F\left( q,Q\right) }{\partial q\partial Q}\right| e^{-iF\left(
q,Q\right) }\Theta _{E}\left( q\right)  \label{inv}
\end{equation}
The canonical transformation defined by $\left( \ref{can}\right) $ does
satisfy the condition $\left( \ref{con}\right) .$ The function $F\left(
q,Q\right) $ coincides then with the generating function $F_{1}\left(
q,Q\right) .$ Once defined this ``canonical quantum transformation'' the
physical solutions of the Wheeler-De Witt equation can be found by
transforming the solutions of the Schr\"{o}dinger equation. Finally the
question of defining proper boundary conditions for the solutions of the
Wheeler-De Witt equation without knowing how to reduce the system will be
addressed.

\section{Application to the Taub model}

\smallskip

\subsection{Desparametrization}

\smallskip

We will study the application of the formalism displayed in the previous
section to the particular case known as Taub model. In Bianchi cosmological
models \cite{ryan} the minisuperspace is a three dimensional manifold
parametrized by two parameters $\left( \beta _{+},\beta _{-}\right) $
measuring the spatial anisotropy and a parameter $\alpha $ measuring the
volume of the Universe (Misner parametrization). The Hamiltonian constraint
for minisuperspace models has the form
\begin{equation}
H=e^{3\alpha }\left\{ -p_{\alpha }^{2}+p_{+}^{2}+p_{-}^{2}+e^{-4\alpha
}\left[ V\left( \beta _{+},\beta _{-}\right) -1\right] \right\}  \label{gh}
\end{equation}
where $\left( p_{\alpha },p_{+},p_{-}\right) $ are the momenta canonically
conjugate to $\left( \alpha ,\beta _{+},\beta _{-}\right) $ and the
potential $V\left( \beta _{+},\beta _{-}\right) $ depends upon the
particular Bianchi model.

The Taub model is a particular case of the Bianchi IX for $\beta _{-}=0,$ $%
p_{-}=0$. For this case the resulting Hamiltonian constraint is
\begin{equation}
H=-p_{\alpha }^{2}+p_{+}^{2}+12\pi ^{2}e^{-4\Omega }(e^{-8\beta
_{+}}-4e^{-2\beta _{+}})  \label{mn}
\end{equation}
scaling the Hamiltonian with the factor $e^{-3\alpha }$ . If we define the
variables $u$ and $v$ by
\begin{eqnarray}
\alpha &=&v-2u  \label{jk} \\
\beta _{+} &=&u-2v  \nonumber
\end{eqnarray}
the resulting Hamiltonian is (multiplied by 1/6)
\begin{equation}
H=\frac{1}{6}\left( p_{v}^{2}+36\pi ^{2}e^{12v}\right) -\frac{1}{6}\left(
p_{u}^{2}+144\pi ^{2}e^{6u}\right)  \label{hami}
\end{equation}

By means of this coordinate transformation we could in fact separate in the
Hamiltonian a subsystem depending on just one pair of canonical variables,
which will work as a clock for the other subsystem. A global time is defined
by means of the canonical transformation defined in $\left( \ref{can}\right)
.$ In our case the new variables are
\begin{eqnarray}
t &=&Arc\sinh \left( -\frac{p_{v}}{6\pi }e^{-6v}\right)  \label{pl} \\
p_{t}^{2} &=&\frac{1}{36}\left( p_{v}^{2}+36\pi ^{2}e^{12v}\right)
\label{nj}
\end{eqnarray}

The generator of this transformation is
\begin{equation}
F_{1}\left( v,t\right) =-\pi e^{6v}\sinh t  \label{se}
\end{equation}

The Hamiltonian in the new variables results to be
\begin{equation}
H=6p_{t}^{2}-\frac{1}{6}\left( p_{u}^{2}+144\pi ^{2}e^{6u}\right)  \label{mp}
\end{equation}

This expression can be factorized in order to obtain a Hamiltonian linear in
$p_{t}$, so giving
\begin{equation}
H=\left( \sqrt{6}p_{t}+\frac{1}{\sqrt{6}}\sqrt{p_{u}^{2}+\pi ^{2}e^{6u}}%
\right) \left( \sqrt{6}p_{t}-\frac{1}{\sqrt{6}}\sqrt{p_{u}^{2}+144\pi
^{2}e^{6u}}\right)  \label{vb}
\end{equation}

The constraint $H=0$ is fulfilled if one of the factors is null on the
constraint surface. The other factor has, on the constraint surface, a
definite sign, so playing the role of the factor $f$ defined before. The
scaled Hamiltonian is
\begin{mathletters}
\[
\tilde{H}=\frac{H}{\sqrt{6}f}=p_{t}+\frac{1}{6}\sqrt{p_{u}^{2}+144\pi
^{2}e^{6u}}=p_{t}+h_{u}
\]
with
\end{mathletters}
\[
f=\left( \sqrt{6}p_{t}-\frac{1}{\sqrt{6}}\sqrt{p_{u}^{2}+144\pi ^{2}e^{6u}}%
\right)
\]

\subsection{Quantization}

\smallskip

In order to quantize the reduced system we will make the substitution $%
p_{t}\rightarrow -i\frac{\partial }{\partial t}$, $p_{u}\rightarrow -i\frac{%
\partial }{\partial u}$ and impose the constraint $\hat{H}\Psi \left(
t,u\right) =0$ yielding the following Schr\"{o}dinger equation
\begin{equation}
i\frac{\partial \Psi \left( t,u\right) }{\partial t}=\hat{h}_{u}\left( u,-i%
\frac{\partial }{\partial u}\right) \Psi \left( t,u\right)  \label{ce}
\end{equation}
Inserting solutions of the form $\Psi \left( t,u\right) =\phi \left(
u\right) e^{-i\sqrt{\frac{E}{6}}t}$ we obtain a modified Bessel equation for
the function $\phi \left( u\right) .$ The solutions of this equation are the
modified Bessel functions
\begin{equation}
\phi \left( u\right) =CK_{2i\sqrt{\varepsilon }}\left( 4\pi e^{3u}\right)
+DI_{2i\sqrt{\varepsilon }}\left( 4\pi e^{3u}\right)  \label{wc}
\end{equation}
The functions $I_{2i\sqrt{\varepsilon }}\left( 4\pi e^{3u}\right) $ should
be discarded because they diverge when $u\rightarrow \infty $ (classically
forbidden zone). The solutions corresponding to the quantization of the
reduced system are therefore
\begin{equation}
\Psi \left( t,u\right) =Ce^{-i\sqrt{\varepsilon }t}K_{2i\sqrt{\varepsilon }%
}\left( 4\pi e^{3u}\right)  \label{my}
\end{equation}
On the other hand the Wheeler-De Witt equation associated with the
Hamiltonian $\left( \ref{hami}\right) $ is
\begin{equation}
\frac{1}{6}\left[ \left( -\frac{\partial ^{2}}{\partial v^{2}}+36\pi
^{2}e^{12v}\right) -\frac{1}{6}\left( -\frac{\partial ^{2}}{\partial u^{2}}%
+144\pi ^{2}e^{6u}\right) \right] \varphi \left( v,u\right) =0  \label{c}
\end{equation}
whose solutions are
\begin{equation}
\varphi \left( v,u\right) =\left[ AK_{i\sqrt{\varepsilon }}\left( \pi
e^{6v}\right) +BI_{i\sqrt{\varepsilon }}\left( \pi e^{6v}\right) \right]
\left[ CK_{2i\sqrt{\varepsilon }}\left( 4\pi e^{3u}\right) +DI_{2i\sqrt{%
\varepsilon }}\left( 4\pi e^{3u}\right) \right]  \label{bi}
\end{equation}

It would not be correct to impose the same kind of boundary conditions used
to discard the functions $I_{2i\sqrt{\varepsilon }}\left( 4\pi e^{3u}\right)
$ in the quantization of the reduced system. The variable $v$ is not a
dynamical variable but the variable associated with the clock of the system.
It is by no means obvious that the physical solutions should go to zero in
the classical forbidden zone.

In order to select the physical solutions we will try to apply the ``quantum
canonical transformations'' defined in $\left( \ref{inv}\right) $ to the
solutions of the Wheeler-De Witt equation. The physical solutions will be
those whose transformed functions are the solutions of the Schr\"{o}dinger
equation $e^{-i\sqrt{\varepsilon }t}.$ We will begin by transforming the
functions which go to zero in the classically forbidden zone, i.e., the
functions $\Theta \left( v\right) =K_{i\sqrt{\varepsilon }}\left( \pi
e^{6v}\right) .$ The transformed functions are
\begin{eqnarray}
\Phi \left( t\right) &=&N\int_{-\infty }^{+\infty }dv6\pi e^{6v}\cosh
te^{i\pi e^{6v}\sinh t}K_{i\sqrt{\varepsilon }}\left( \pi e^{6v}\right)
\nonumber \\
\ &=&\frac{\pi N}{4\sinh \left( \frac{\pi \sqrt{\varepsilon }}{2}\right)
\cosh \left( \frac{\pi \sqrt{\varepsilon }}{2}\right) }\left[ e^{\frac{\pi
\sqrt{\varepsilon }}{2}}e^{i\sqrt{\varepsilon }t}-e^{-\frac{\pi \sqrt{%
\varepsilon }}{2}}e^{-i\sqrt{\varepsilon }t}\right]  \label{pupi}
\end{eqnarray}
In this way it is manifest that the transformation of the functions $\Theta
\left( v\right) =K_{i\sqrt{\varepsilon }}\left( \pi e^{6v}\right) $ do not
give the solutions of the Schr\"{o}dinger equation $e^{-i\sqrt{\varepsilon }%
t}.$ On the contrary they correspond to a combination of positive and
negative energies states
\begin{equation}
K_{i\sqrt{\varepsilon }}\left( \pi e^{6v}\right) \leftrightarrow \frac{\pi N%
}{4\sinh \left( \frac{\pi \sqrt{\varepsilon }}{2}\right) \cosh \left( \frac{%
\pi \sqrt{\varepsilon }}{2}\right) }\left[ e^{\frac{\pi \sqrt{\varepsilon }}{%
2}}e^{i\sqrt{\varepsilon }t}-e^{-\frac{\pi \sqrt{\varepsilon }}{2}}e^{-i%
\sqrt{\varepsilon }t}\right]  \label{uuu}
\end{equation}
By transforming the right side of $\left( \ref{uuu}\right) $ one should
recover the original function $\Theta \left( v\right) =K_{i\sqrt{\varepsilon
}}\left( \pi e^{6v}\right) ,$ so one obtains the factor $N=\frac{1}{\sqrt{%
\pi }},$ which does not depend on the energy.

As the functions $\Theta \left( v\right) =K_{i\sqrt{\varepsilon }}\left( \pi
e^{6v}\right) $ are not definite energy states, we will apply the
transformation $\left( \ref{inv}\right) $ to the other subspace of
solutions, i.e., to the functions $I_{\pm i\sqrt{\varepsilon }}\left( \pi
e^{6v}\right) ,$ which diverge in the classically forbidden zone$.$ The
resulting integral has the form
\begin{equation}
\Phi \left( t\right) =\frac{1}{\sqrt{\pi }}\int_{-\infty }^{+\infty }dv6\pi
e^{6v}\cosh te^{i\pi e^{6v}\sinh t}I_{\pm i\sqrt{\varepsilon }}\left( \pi
e^{6v}\right)  \label{rt}
\end{equation}
This integral diverges unless one gives an imaginary part $\eta $ to $t.$
Replacing $t\rightarrow t+i\frac{\pi }{2}$ in $\left( \ref{rt}\right) $ one
can actually perform the integration. Replacing $t\rightarrow t-i\frac{\pi }{%
2}$ in the result of the integral one can go back to the original real time
variable, obtaining the correspondence
\begin{equation}
I_{\pm i\sqrt{\varepsilon }}\left( \pi e^{6v}\right) \leftrightarrow \Phi
\left( t\right) =\frac{i}{\sqrt{\pi }}e^{\mp \sqrt{\varepsilon }\frac{\pi }{2%
}}e^{\mp i\sqrt{\varepsilon }t}  \label{pup}
\end{equation}
The functions $I_{i\sqrt{\varepsilon }}\left( \pi e^{6v}\right) $ and $I_{-i%
\sqrt{\varepsilon }}\left( \pi e^{6v}\right) $ do represent then the
positive and negative energy states respectively \footnote{%
This result can be verified by testing the consistence of $\left( \ref{pup}%
\right) $ with $\left( \ref{uuu}\right) $ and the expression $\cite{abram}$
\begin{equation}
K_{v}\left( z\right) =\frac{\pi }{2}\frac{I_{-v}\left( z\right) -I_{v}\left(
z\right) }{\sin \left( v\pi \right) }  \label{qwe}
\end{equation}
In fact, transforming the right side of $\left( \ref{qwe}\right) $ using $%
\left( \ref{pup}\right) $ one does obtain the right side of $\left( \ref{uuu}%
\right) ,$ verifying in this way the coherence of the found correspondences $%
\left( \ref{uuu},\text{ }\ref{pup}\right) $ between both representations$.$}%
. In this way we could establish which subspace of the whole space of
solutions of the Wheeler-De Witt equation is the physical one. It is
remarkable that the functions in the selected subspace do not decay in the
classically forbidden zone.

\smallskip

\subsection{Boundary conditions}

\smallskip

It would be interesting if one could define certain boundary conditions
which would not rely on the fact that one knows how to reduce the system,
and which would select the same physical functions imposed by the
quantization of the reduced system. Let us start by considering the
following Hamiltonian
\begin{equation}
H=p_{1}^{2}+V\left( q_{1}\right) -h_{q_{\mu }}\left( q_{\mu },p_{\mu
}\right) \text{, }\mu =2,..,N  \label{ham}
\end{equation}
and suppose that $\varphi \left( q_{1},q_{\mu }\right) =$ $\Theta \left(
q_{1}\right) \phi \left( q_{\mu }\right) $ is the solution of the Wheeler-De
Witt equation associated with this Hamiltonian. Instead of performing a
canonical transformation so that the new Hamiltonian is quadratic in the new
momentum $p_{t}$, we will study the wave functions obtained by solving the
Wheeler-De Witt equation associated with the Hamiltonian $\left( \ref{ham}%
\right) $ in the region $L$ where $V\left( q_{1}\right) $ tends to zero. In
this region the Hamiltonian is
\begin{equation}
H=p_{1}^{2}-h_{q_{\mu }}\left( q_{\mu },p_{\mu }\right)  \label{buy}
\end{equation}

The solutions of the quantum-mechanical system corresponding to the sheet in
which $p_{1}$ is negative (positive energy solutions) will be combinations
of $\varphi \left( q_{1},q_{\mu }\right) =\phi \left( q_{\mu }\right) e^{-i%
\sqrt{\varepsilon }q_{1}}.$ One would expect that these solutions do
coincide with the asymptotic expressions in the region $L$ of the functions $%
\varphi \left( q_{1},q_{\mu }\right) =$ $\Theta \left( q_{1}\right) \phi
\left( q_{\mu }\right) $, i.e., it would be necessary that
\begin{eqnarray}
\Theta \left( q_{1}\right) &\rightarrow &e^{-i\sqrt{\varepsilon }q_{1}}
\label{ect} \\
q_{1} &\rightarrow &L  \nonumber
\end{eqnarray}

The definite energy solutions will thus be those functions $\Theta \left(
q_{1}\right) $ which do behave in the asymptotic region like a plane wave,
ingoing $\emph{or}$ outgoing. This criterium relies on the fact that in the
asymptotic region time is (modulus a sign) the variable $q_{1}.$

In order to test this criterium in the case of the Taub model let us see the
behavior of the Bessel functions in the region where $v\rightarrow -\infty $
$\left( V\left( q_{1}\right) \rightarrow 0\right) $. The asymptotic
expressions are
\begin{eqnarray}
I_{v}\left( z\right) &\sim &\frac{\left( \frac{1}{2}z\right) ^{v}}{\Gamma
\left( v+1\right) }  \label{ecr} \\
K_{v}\left( z\right) &\sim &\frac{\pi }{2\sin \left( v\pi \right) }\left[
\frac{\left( \frac{1}{2}z\right) ^{-v}}{\Gamma \left( -v+1\right) }-\frac{%
\left( \frac{1}{2}z\right) ^{v}}{\Gamma \left( v+1\right) }\right]
\label{iny}
\end{eqnarray}

The asymptotic expression for $K_{v}\left( z\right) $ was obtained using the
formula $\left( \ref{qwe}\right) $. The asymptotic expressions are thus
\begin{equation}
I_{i\sqrt{\varepsilon }}\left( \pi e^{6v}\right) \sim \frac{\left( 36\pi
^{2}\right) ^{\frac{i\sqrt{\varepsilon }}{2}}}{\left( 12\right) ^{i\sqrt{%
\varepsilon }}\Gamma \left( i\sqrt{\varepsilon }+1\right) }e^{i\sqrt{%
\varepsilon }6v}  \label{wxr}
\end{equation}
\begin{equation}
 K_{i\sqrt{\varepsilon }}\left( \pi e^{6v}\right) \sim
\frac{\pi }{2\sin \left( v\pi \right) } \left[ \frac{\left( 36\pi ^{2}\right)
^{-\frac{i\sqrt{\varepsilon }}{2}}}{\left(
12\right) ^{-i\sqrt{\varepsilon }}\Gamma \left( -i\sqrt{\varepsilon }%
+1\right) }e^{-i\sqrt{\varepsilon }6v} \\ -\frac{\left( 36\pi
^{2}\right)^{\frac{i\sqrt{\varepsilon }}{2}}}{\left( 12\right) ^{i\sqrt{\varepsilon
}}\Gamma \left( i\sqrt{\varepsilon }+1\right) }e^{i\sqrt{\varepsilon }6v} \right]
\label{excg}
\end{equation}

In this way we confirm that the functions $K_{i\sqrt{\varepsilon }}\left(
\pi e^{6v}\right) $ do correspond to a combination of positive and negative
energy states. The functions $I_{\pm i\sqrt{\varepsilon }}\left( \pi
e^{6v}\right) $ do correspond to states of positive or negative energy
respectively. The proposed criterium establishes boundary conditions with
the definite meaning of selecting the positive energy states. The space of
solutions of these positive energy states can be endowed with the positive
definite Schr\"{o}dinger inner product. It is remarkable that the proposed
boundary conditions coincides with the criterium proposed by Wald \cite{wald,wald2}%
.

\smallskip

\section{Conclusions}

\smallskip

In this work it was addressed the question of quantizing minisuperspace models by studying
the particular case known as Taub Model. The main two problems which arise in the
canonical approach are the boundary conditions to be imposed on the solution of the
Wheeler-De Witt equation and the inner product to be defined in the corresponding Hilbert
space. In the Taub model it is possible to perform a coordinate transformation in order to
separate in the Hamiltonian constraint a subsystem depending on just one pair of canonical
variables. In this new set of variables the Hamiltonian has thus the form
$H=h_{q_{1}}\left( q_{1},p_{1}\right) -h_{q_{\ \mu }}\left( q_{\mu },p_{\mu }\right) $
where the subsystem $h_{q_{1}}$ will work as the clock of the model. Performing a
canonical transformation it is possible to transform the subsystem $h_{q_{1}}$ in a free
system $h_{t}=p_{t}^{2}.$ This kind of time variables are known as $``$extrinsic time''
because they are associated not only with the coordinates but also with the momenta \cite
{canada,york}. Extrinsic times are specially important in quantum gravity because it is
not possible to reduce the system by identifying an intrinsic time, i.e., a global time
variable in the configuration space, as
it happens for the relativistic particle. The new Hamiltonian $H=$ $%
p_{t}^{2}-h_{q_{\ \mu }}\left( q_{\mu },p_{\mu }\right) $ can be factorized
in two disconnected sheets. In order to satisfy the constraint $H=0$ it is
necessary that one of this factors goes to zero on the constraint surface.
As the other one has thus a definite sign on the constraint surface, it is
possible to scale the Hamiltonian constraint in order to find a Hamiltonian
linear in the new momentum $p_{t}.$ This Hamiltonian can be quantized by
means of an ordinary Schr\"{o}dinger equation. The canonical transformation
used to reduce the system satisfies the necessary conditions to define the
integral transforms which relate the wave functions corresponding to the
quantization of both Hamiltonian systems. It is thus possible to transform
the positive energy solutions of the parabolic Schr\"{o}dinger equation in
order to find out those solutions of the hyperbolic Wheeler-De Witt equation
which are the physical ones. Armed with the knowledge of the physical
solutions, we tried define a criterium to select those solutions without
using the fact that the reduced system is known. In order to do that it was
studied the asymptotic behavior in the free zone $\left( V\left(
q_{1}\right) \rightarrow 0\right) $ of the solutions of the Wheeler-De Witt
equation. We argue that in that area the time is (minus a sign) the variable
$q_{1}$. It is thus necessary that the functions $\Phi \left( q_{1}\right) $
behave as an outgoing $\emph{or}$ ingoing plane wave, i.e., as definite
energies states. The wave functions satisfying this criterium do coincide
with the physical functions selected by reducing the system.

\acknowledgments

This work was supported by Fundaci\'{o}n Antorchas, Consejo Nacional de
Investigaci\'{o}n Cient\'{i}ficas y T\'{e}cnicas (CONICET) and Universidad
de Buenos Aires (Proy. TX 64).



\begin{references}

\bibitem{canada}  \smallskip K. V. Kuchar, \emph{Time and interpretations of
Quantum Gravity, }in Proceedings of the 4$^{th}$ Canadian Conference on
General Relativity and Relativistic Astrophysics, ed. G. Kunstatter, D.
Vincent, J. Williams, World Scientific, Singapore $\left( 1992\right) $.

\bibitem{isham}  J. Butterfield and C. J. Isham, \emph{On the Emergence of
Time in Quantum Gravity}, in The Arguments of Time, ed. J. Butterfield,
Oxford University Press, $\left( 1999\right) .$

\bibitem{extrinseco}  R. Ferraro, \emph{The Problem of Time in Quantum
Gravity}, in Gravitation \& Cosmology, 5, 195, $\left( 1999\right) .$

\bibitem{lanczos}  C. Lanczos, \emph{The variational principles of Mechanics}%
, Dover, N. Y. (1986).

\bibitem{tiempo}  P. H\'{a}j\'{\i }cek, Phys. Rev. D 34, 1040, $\left(
1986\right) .$

\bibitem{gravitation}  C. Misner, K. Thorne and J. A. Wheeler, \emph{%
Gravitation}, Freeman, S. F. $\left( 1973\right) $.

\bibitem{killing}  K. V. Kuchar, \emph{Canonical methods of quantization},
in Quantum Gravity 2: A Second Oxford Symposium, ed. C. J. Isham, R. Penrose
y D. W. Sciama, Clarendon Press, Oxford $\left( 1981\right) $.

\bibitem{beluardi}  S.C. Beluardi and R. Ferraro, Phys. Rev. D 52, 1963,
(1995).

\bibitem{gandur}  G. I. Ghandour, Phys. Rev. D 35, 1289, $\left( 1987\right)
.$

\bibitem{ryan}  M. P. Ryan, Jr. and L. C. Shepley, \emph{Homogeneous
Relativistic Cosmologies}, Princeton University Press, New Jersey (1975).

\bibitem{abram}  \emph{\ Handbook of mathematical functions}, ed. M.
Abramowitz y I. A. Stegun.

\bibitem{wald}  R.M. Wald, Phys. Rev. D 48, 2377, (1993).

\bibitem{wald2}  R.M. Wald and A. Higuchi, Phys. Rev. D 51, 544, (1995).

\bibitem{york}  J. W. York, Phys. Rev. Lett. 28, 1082 (1972).
\end{references}
\end{document}